\documentclass[floatfix,10pt,reprint,aps,prl]{revtex4-1}

\usepackage{graphicx}
\usepackage{bm}

\begin{document}

\author{Florian Pielmeier}
 \email{florian1.pielmeier@physik.uni-regensburg.de}
 \affiliation{Institute of Experimental and Applied Physics, University of Regensburg, D-93053 Regensburg, Germany}

\author{Franz J. Giessibl}%

\affiliation{Institute of Experimental and Applied Physics, University of Regensburg, D-93053 Regensburg, Germany}

\date{\today}

\title{\textbf{Spin Resolution and Evidence for Superexchange on NiO(001) observed by Force Microscopy}}

\begin{abstract}

The spin order of the nickel oxide (001) surface is resolved, employing non-contact atomic force microscopy at 4.4 K using bulk Fe- and SmCo-tips mounted on a qPlus sensor that oscillates at sub-50\,pm amplitudes. The spin-dependent signal is hardly detectable with Fe-tips. In contrast, SmCo-tips yield a height contrast of $1.35\, \mathrm{pm}$ for Ni ions with opposite spins. SmCo tips even show a small height contrast on the O atoms of $0.5\, \mathrm{pm}$ within the 2x1 spin unit cell, pointing to the observation 
of superexchange. We attribute the increased signal-to-noise ratio to the increased magnetocrystalline anisotropy energy of SmCo, which stabilizes the magnetic moment at the apex. Atomic force spectroscopy on the $\mathrm{Ni \uparrow}$, $\mathrm{Ni \downarrow}$ and O lattice site reveals a magnitude of the exchange energy of merely
$1\,\mathrm{meV}$ at the closest accessible distance with an exponential decay length of $\lambda_{exc}=18$\,pm.
\end{abstract}
\maketitle

High resolution non-contact Atomic Force Microscopy (nc-AFM) detects short-range chemical interactions between the foremost tip atoms and sample atoms, enabling atomic resolution imaging and quantitative force measurements \cite{Binnig1986,Giessibl1995,Lantz2001}. By equipping an AFM with a magnetic probe tip, the sample magnetization can be studied \cite{Martin1987} at a resolution
of several tens of nanometers \cite{Koblischka2003}. Wiesendanger et al. estimated in 1990, that magnetic exchange interactions that occur in spin-polarized scanning tunneling microscopy can
amount to about one pN per \AA$^2$ of tip area \cite{Wiesendanger1991}. Several calculations predicted even larger magnitudes of exchange forces \cite{Mukasa1994a,Ness1995,Nakamura1997,Foster2001,Momida2005}. Once atomic resolution by AFM in ultrahigh vacuum (UHV) became feasible, extended efforts to detect exchange interactions by nc-AFM on NiO at $T=4$\,K and 300\,K were conducted \cite{Hosoi2000,Allers2001,Hosoi2004,Schmid2008}, initially without success.
In 2007, Kaiser et al. proved the feasibility of Magnetic Exchange Force Microscopy (MExFM) by imaging the $(2\times1)$ spin pattern on the antiferromagnetic insulator NiO \cite{Kaiser2007a}.
The experiment was conducted at liquid helium temperatures, using an iron coated silicon cantilever where the magnetization of the tip was stabilized by applying a 5\,T magnetic field  \cite{Kaiser2007a,Kaiser2008,Schwarz2009}.
The exchange interaction between tip and sample is qualitatively described by the Heisenberg model, $H=-J_{12} \vec{S_1}\cdot \vec{S_2}$, where $J_{12}$ is the exchange coupling constant. For $3d$ transition metals a large magnetic moment of the foremost tip atom is desirable for achieving a high signal-to-noise-ratio \cite{Schwarz2009}.

In this Letter, we report on the detection of spin contrast on the NiO(001) surface without applying an  external magnetic field. We analyse the dependence of the contrast for Fe- and SmCo-tips. Both tips reveal the antiferromagnetic structure of NiO(001), but SmCo-tips yield a 3-10 times higher spin contrast than Fe-tips. With the magnetic moments of $\mu_{Fe}=2.2\,\mu_B$, $\mu_{Co}=1.7\,\mu_B$ and $\mu_{Sm}=0.4\,\mu_B$  \cite{Givord1979}, this finding shows that $\mu$ is not the only parameter that determines spin contrast in MExFM.
We attribute the increased contrast in case of SmCo-tips to the  higher magnetocrystalline anisotropy energy (MAE) compared to Fe, which stabilizes the spin orientation of the front atom. Furthermore we present $\Delta f(z)$-curves acquired with a SmCo-tip and evaluate the magnitude of the exchange interaction on NiO. We find that its magnitude is only about 1/50 of the exchange interaction
between Fe-tips and an antiferromagnetically ordered Fe monolayer on W(001) \cite{Schmidt2011}.

 \begin{figure}
	\centering
		\includegraphics[width=0.48\textwidth]{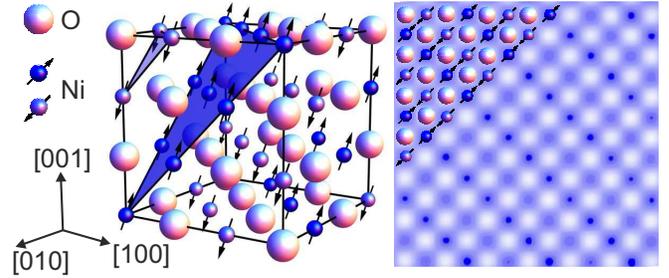}
	\caption{(Color online). Left: crystal structure and magnetic structure of nickel oxide (see text). Right: Slightly low pass filtered \cite{Horcas2007} MExFM topography image of NiO(001), showing the $(2\times 1)$ unit cell of the surface. Imaging parameters: SmCo-tip, $k = 2425\,\textnormal{N/m}$, $f_0 = 39.761\,\textnormal{kHz}$, $A = 36\,\textnormal{pm}$, $Q = 31,000$ and bias voltage $U_{bias} = 0.06\,\textnormal{V}$.}
	\label{fig:niokristall}
\end{figure}

Forces are measured by frequency modulation atomic force microscopy \cite{Albrecht1991}, where the force sensor with stiffness  $k$ , eigenfrequency $f_0$ and quality factor $Q$ oscillates at a constant amplitude $A$ and is subject to a frequency shift $\Delta f=f-f_0$ that is directly related to the averaged tip-sample force gradient via $\left\langle k_{ts} \right\rangle=\frac{2k}{f_0}\Delta f$ \cite{Giessibl1997}. Forces have been derived by deconvolving the frequency shift $\Delta f$ with the Sader-Jarvis-method \cite{Sader2004}.
Optimal sensitivity to short-range forces is ensured by operating the qPlus force sensor at amplitudes below $100\,\textnormal{pm}$ \cite{Giessibl1999,Ternes2008,Gross2009}.
The sensor can be equipped with any tip material, in a previous study on NiO, cobalt was used due to its lower chemical reactivity \cite{Giessibl1998,Schmid2008}.
Iron tips were electrochemically etched from a high purity iron wire ($99.99+\%$), whereas a sharp piece of a SmCo permanent magnet was glued to the qPlus sensor to obtain a SmCo-tip \cite{Herz2003}. Before the tips where introduced into the UHV system, they were sharpened by focussed ion beam (FIB) etching.
The native oxide layer of bulk metal tips is removed by field evaporation \cite{Muller} in UHV, afterwards the sensors are transfered \textit{in-situ} to the microscope within 15 minutes.
The measurements were carried out on an Omicron LT/qPlus system in UHV ($p\leq10^{-10}\,\textnormal{mbar}$) and at a temperature of $4.4\,\textnormal{K}$.

\begin{figure}
	\centering
		\includegraphics[width=0.44\textwidth]{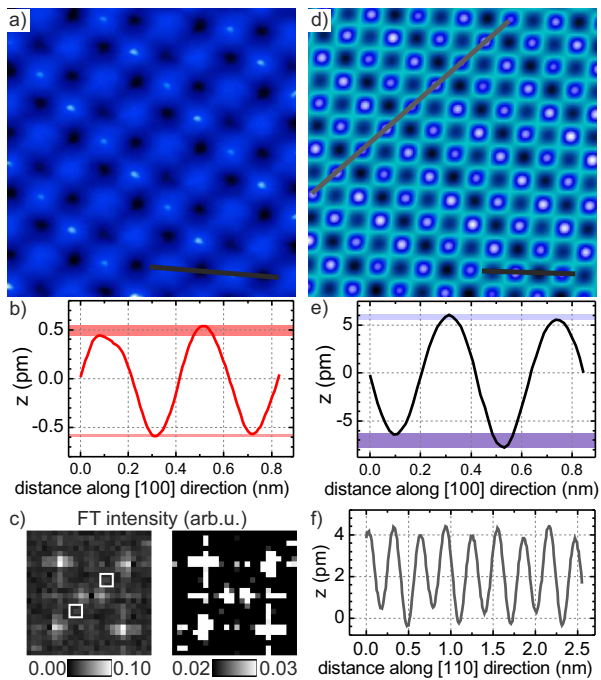}
	\caption{ (Color online). MExFM data acquired with Fe (left) and SmCo (right) tips. a) Low-pass filtered ($2\, \times\, 2$) unit cell averaged topography image $(2 \times 2 \,\mathrm{nm^2})$ showing the row-wise contrast, for image processing details see \cite{supp}. Line profile in b) shows a height difference between the local maxima of $0.1\,\textnormal{pm}$, the average atomic corrugation is $1.1\,\textnormal{pm}$.
	 c) Fourier spectra of the raw data corresponding to a), in normal and high contrast (right). 	
	 d) Low-pass filtered topography data $(2.7 \times 2.7\, \mathrm{nm^2})$ acquired with a SmCo-tip. Each second Ni row appears darker. e) Line profile, revealing a difference between Ni sites of up to $1.35 \,\textnormal{pm}$. The height of the oxygen sites within one magnetic unit cell varies by $0.5\,\textnormal{pm}$.
	 f) Line profile showing the periodicity of the height variations on oxygen sites.
	Parameters for Fe- (SmCo-) sensor: $k = 1800\,\textnormal{N/m}$ ($2425\,\textnormal{N/m}$), $f_0 = 59.369\,\textnormal{kHz}$ ($39.761\,\textnormal{kHz}$), $A = 50\,\textnormal{pm}$ ($36\,\textnormal{pm}$), $Q = 1,362,000$ ($31,000$) and $U_{bias} = 6.8\,\textnormal{V}$ ($0.06\,\textnormal{V}$).	
	 }
	\label{fig2}
\end{figure}

The structure of the antiferromagnetic insulator nickel oxide is shown in Fig. \ref{fig:niokristall}. NiO exhibits a rock salt structure with a lattice constant of $a=417\,\textnormal{pm}$.
Nickel atoms in $\{111\}$ planes are coupled ferromagnetically and neighbouring Ni planes are coupled antiferromagnetically via superexchange mediated by the oxygen atoms. This leads to an antiferromagnetic structure at the $(001)$ surface with alternating spin orientations of nickel atoms along the $\langle110\rangle$ direction.
The NiO crystal (SurfaceNet, Rheine, Germany) was cleaved in-situ to obtain clean and flat terraces up to $100\,\textnormal{nm}$ in width. Cleaved NiO surfaces exhibit a bulk-terminated orientation of magnetic moments \cite{Hillebrecht2001}. On the right in Fig. \ref{fig:niokristall} a model of the surface atomic and magnetic structure is superimposed to a high-resolution MExFM image aquired with a SmCo-tip, showing alternating rows of oppositely aligned Ni atoms along the $[1\bar{1}0]$ direction. When imaging with a metallic tip, O atoms usually appear as maxima in constant frequency shift mode \cite{Foster2001}, and the minima refer to Ni sites. The difference in apparent height between the two nickel sites is due to the exchange interaction which adds to the chemical interaction depending on the spin alignment of the surface Ni atoms relative to the tip moment. A direct exchange mechanism has been predicted for an Fe atom probing the NiO surface \cite{Momida2005}.

As in all successful MExFM experiments on NiO so far \cite{Kaiser2007a,Kaiser2008,Schwarz2009}, we used Fe-tips in our initial experiments.
Here, we measure exchange contrast on NiO using Fe tips \emph{without} an external magnetic field, yielding a very weak exchange contrast
that extends over a narrow distance range of about $10-20$\,pm \cite{supp}.
The small width of the distance range where exchange forces are detectable indicates that the stability of the spin orientation of the tip apex atom is easily altered by increasing tip-sample interaction forces. Locally, the stability of the spin orientation is governed by the directional dependent magnetocrystalline anisotropy (MA). Hence, the tip cluster orientation my effect the ontrast in MExFM experiments. The magnetic easy axis of bulk bcc-iron is parallel to $\langle100\rangle$ directions \cite{Cullity}. As a next step we use a tip with a known tip cluster orientation, achieved by probing the tip apex with a CO molecule adsorbed on Cu(111) \cite{Welker2012}. As both Fe and W are bcc materials, we observe the same symmetries for Fe tips \cite{supp} as we did for W tips in  \cite{Welker2012}. After the Fe-tip was characterized by the CO-method, the Cu sample is removed and the cleaved NiO sample is introduced into the microscope. After carefully approaching the NiO(001) surface the metallic nature of the tip apex was confirmed by $\Delta f(U)$ curves, where the absence of charging effects or tunneling to localized states is an indication for a metallic tip apex \cite{supp,Teobaldi2011}. Electrostatic forces were minimized by applying a bias voltage to the sample.

Figure \ref{fig2}(a) shows a low-pass filtered, unit cell averaged topographic image acquired with a Fe-tip, which is oriented along a $\langle100\rangle$ direction \cite{supp}. The image was acquired in constant height mode and the frequency shift $(\Delta f)$ was converted to topography, see \cite{supp}. A $2 \times 2$ unit cell was used to avoid superimposing the data with the expected $2 \times 1$ magnetic unit cell. The additional modulation of the atomic contrast can be identified, as a row-wise changing apparent height of the \textit{maxima}. The topography line profile in \ref{fig2}(b) shows a difference between two local maxima of only $0.1\,\mathrm{pm}$, the average atomic corrugation is $1.1\,\mathrm{pm}$. In Fig. \ref{fig2}(c), two Fourier spectra of the unfiltered raw data corresponding to a) are shown. Two additional peaks (solid white boxes) appear at half the inverse lattice vector along a line from the lower left to the upper right corner. There are two possible reasons for the appearance of larger spin modulation on top of the maxima compared to minima, either the Ni sites are imaged as maxima, or due to superexchange on O sites which might be stronger in this distance regime.

Although the spin contrast using an oriented Fe-tip is larger on maxima than on minima in Fig. 2 a), the magnitude of the spin contrast is in good agreement with our initial experiments with uncharacterized iron tips, where it reached up to $0.4\,\mathrm{pm}$ on top of a small chemical interaction causing $1.6\,\mathrm{pm}$ corrugation (Figs. 1, 2 in \cite{supp}). MExFM with Fe-tips only yields a weak spin contrast over a thin distance range where chemical forces are small and the spin-dependent signal is lost when the tip height deviates from the ideal height by more than $\pm15\,\mathrm{pm}$. Even though the observation of low spin contrast can be due to an unfavorable alignment of tip and sample spins, Fe tips systematically yield low spin contrast as we performed several experiments with different Fe-tips and investigated different spots of a given NiO sample.
The instability of the spin orientation of the apex atoms upon increased chemical bonding forces between tip and sample indicates that the spin orientation of the apex atoms rotates at closer distances to maximize the chemical interaction and that the MA in Fe is not high enough to stabilize the magnetic moment of the front atom. Indeed, the magnetocrystalline anisotropy energy (MAE) for bcc iron is only $2.4\, \mathrm{\mu eV/atom}$, whereas hcp Co already has a MAE of $45 \,\mathrm{\mu eV/atom}$ \cite{Cullity}. Materials with even higher MAEs are permanent magnets like samarium-cobalt alloys, their MAE is about 20-40 times larger than hcp Co and hence about a factor of 500 higher than the MAE of bulk bcc iron \cite{Weller2000,Burkert2004}. Using such high MAE materials as tips in MExFM experiments should lead to a higher stability of the spin orientation of the tip apex. To test this hypothesis, the MExFM measurements on NiO were repeated with bulk SmCo-tips. The results are shown in Fig. \ref{fig2}(d)-(f), the additional modulation is clearly apparent in the low-pass filtered topography image d) of the NiO(001) surface. A line profile from the low pass filtered image is displayed in e), the average atomic corrugation is $12.9 \,\mathrm{pm}$. The difference between the two local minima due to exchange interaction is $1.35\,\mathrm{pm}$ (dark blue shaded bar).  The chemical and spin resolution is independent of the scan direction \cite{supp}.

Interestingly, a small height difference of $0.5\,\mathrm{pm}$ (light blue shaded bar) between the oxygen sites (local maxima) can be identified. These height variations show the same periodicity as the height variation on Ni sites, Fig. \ref{fig2}(f).
An additional modulation on top of the oxygen atoms has already been discussed in \cite{Kaiser2008}. There, it was attributed to a magnetic double tip, mainly because the line profile showed an asymmetric, wedgelike shape of the atoms. Furthermore a direct exchange mechanism between the magnetic moment of the oxygen and the tip moment is unlikely as it is about an order of magnitude smaller than the moment on the nickel sites \cite{Ködd2002,Momida2005}. As the line profile in Fig. \ref{fig2}(e) has an overall sinusoidal shape, we believe that the height difference on top of the oxygen sites is not due to a magnetic double tip but rather caused by an indirect exchange mechanism between the tip moment and the second layer nickel atoms underneath the oxygen.

\begin{figure*}
	\centering
		\includegraphics[width=0.98\textwidth]{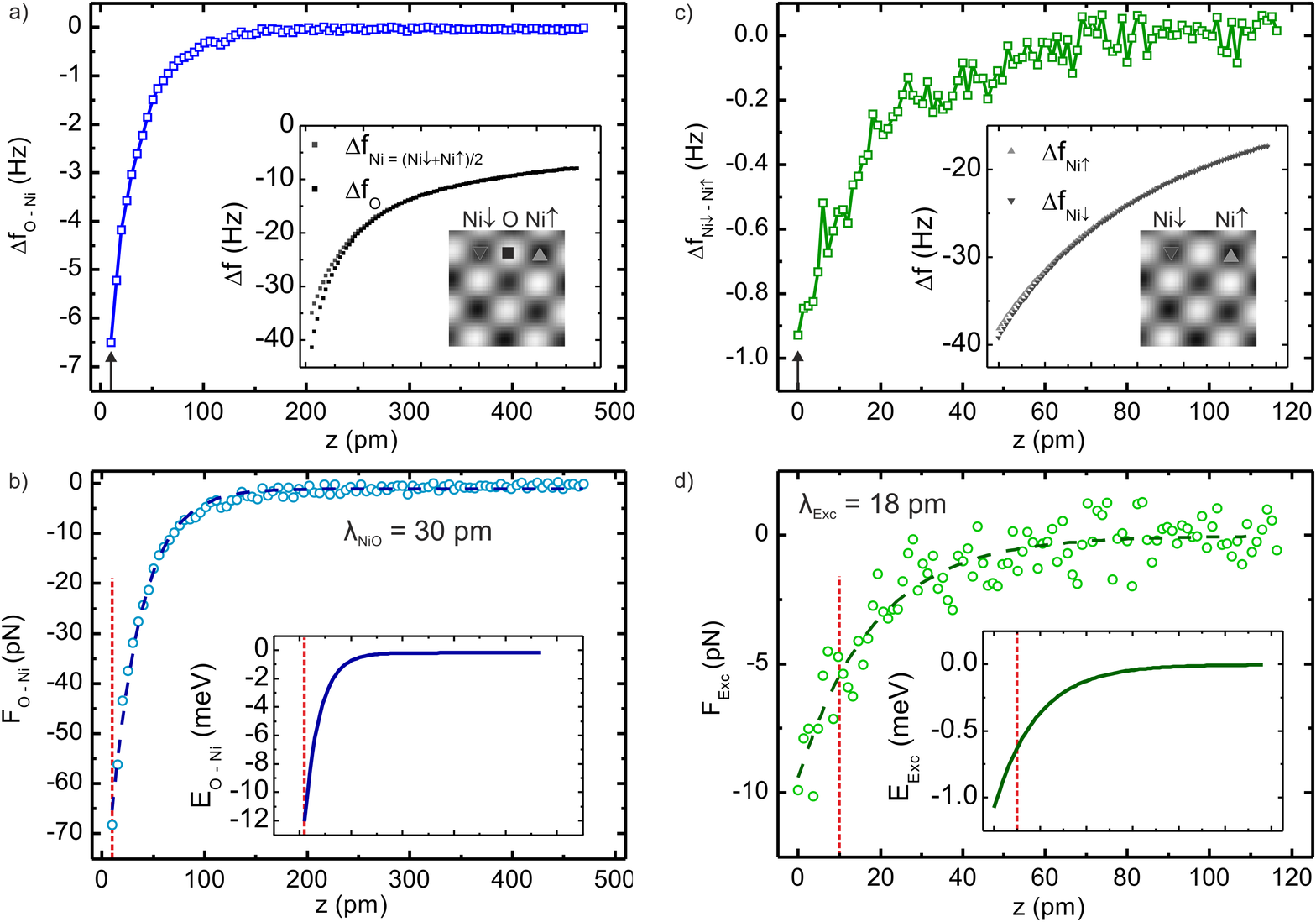}
	\caption{(Color online). a) $\Delta f(z)$-spectra on $\mathrm{Ni}$ and O sites (inset) and difference between the curves (blue). Starting positions of spectra are indicated by arrows in a) and c), while the dashed red lines in b) and d) indicate the distance $z=10\,\mathrm{pm}$ where Fig. 2d) was imaged. b) Force difference between the tip and $\mathrm{Ni}$/O sites, including an exponential fit $\propto \exp{(-z/\lambda)}$ with  $\lambda = 30\,\textnormal{pm}$ (dashed blue line). Integration of the forces yields the energy difference between Ni and O sites, which reaches $12\,\textnormal{meV}$ (inset).
	 c) $\Delta f(z)$-curves on $\mathrm{Ni\downarrow}$ and $\mathrm{Ni\uparrow}$ sites (inset) and difference (green). The resulting exchange force $F_{Exc}$ in d) has an even smaller decay length than the chemical interaction of only $\lambda_{\mathrm{Exc}}=18\,\mathrm{pm}$.}
	\label{fig3}
\end{figure*}

To evaluate the distance dependence of the atomic and exchange interactions, $\Delta f(z)$ curves with the SmCo-tip from Fig. \ref{fig2}(d) were acquired on three different sites, which are marked in the insets of figures \ref{fig3}a) and c). Namely, O and the two different Ni sites, which are denoted as $\mathrm{Ni\downarrow}$ and $\mathrm{Ni\uparrow}$ for the following discussion. The value of $z=0$ indicates the point of closest approach in the $\Delta f(z)$ curves in Fig. \ref{fig3}(c), whereas the curves in Fig. \ref{fig3}(a) start at $z=10\,\mathrm{pm}$. The image in Fig. \ref{fig2}(d) was also acquired at $z=10$\,pm, marked by the dashed red lines in Fig. \ref{fig3}b)+d). The difference in frequency shift between the O and the average of the Ni sites $\Delta f_{\mathrm{O-Ni}}=\Delta f_{\mathrm{O}}-\Delta f_{\mathrm{Ni = (Ni\downarrow-Ni\uparrow})/2}= 6.5\,\mathrm{Hz}$  at $z=10\,\textnormal{pm}$ [\ref{fig3}(a)]. Open circles in Fig. \ref{fig3}(b) depict the corresponding force values. Fitting an exponentially decaying function we obtain a value of $F_{\mathrm{O-Ni}}=-65\,\textnormal{pN}$ at the imaging distance and a decay length $\lambda_{\mathrm{NiO}}=30\,\mathrm{pm}$.
The difference between  $\mathrm{Ni\downarrow}$ and $\mathrm{Ni\uparrow}$ at $z=10$\,pm is $\Delta f_{\mathrm{Ni\downarrow-Ni\uparrow}}=\Delta f_{\mathrm{Ni\downarrow}}-\Delta f_{\mathrm{Ni\uparrow}}=0.93\,\mathrm{Hz} $ [Fig. \ref{fig3}(c)]. As the Ni sites are chemically equivalent, the difference is purely due to short range magnetic exchange interactions. The exchange force is shown in Fig. \ref{fig3}(d), indicating $F_{\mathrm{Ni\downarrow-Ni\uparrow}}=-5.4\,\mathrm{pN}$ at $z=10$\,pm, and a decay length of $\lambda_{\mathrm{Exc}}=18\,\mathrm{pm}$. The difference between chemical and exchange interaction on NiO with SmCo-tips is given by the ratio of $F_{\mathrm{O-Ni}}/F_{\mathrm{Ni\downarrow-Ni\uparrow}}=-65\,\mathrm{pN}/-5.4\,\mathrm{pN}=12$. Due to the different decay lengths $\lambda_{\mathrm{NiO}}$ and $\lambda_{\mathrm{Exc}}$ for the chemical and exchange interactions the difference in energy is even larger, obtaining a factor of $E_{\mathrm{O-Ni}}/E_{\mathrm{Ni\downarrow-Ni\uparrow}}=-12\,\mathrm{meV}/-0.6\,\mathrm{meV}=20$. Obviously, the main challenge in obtaining spin resolution on NiO is to discriminate the exchange from the chemical interactions.
Theoretical predictions, where an Fe atom probes the NiO surface, find values of the chemical forces in the range of $\mathrm{nN}$ and exchange forces on the order of $0.1\,\mathrm{nN}$ \cite{Momida2005}. The experimental exchange force on NiO(001) is about $10\,\mathrm{pN}$, an order of magnitude smaller and even the chemical forces are below $100\,\mathrm{pN}$ in the experimental distance range. Note, although the SmCo data is not directly comparable with these Fe calculations, the smaller contrast we found for Fe-tips implies that the exchange forces are even smaller in this case. As the exchange force and energy decrease monotonically with decreasing tip-sample distance, there is no indication for a change in the magnetic coupling, as predicted for Fe-tips, within the resolution of our measurements \cite{Momida2005}. NiO is a strongly correlated electron system, which makes it in general challenging for \textit{ab-initio} calculations. Therefore our measurement of the short range exchange interaction on NiO(001) can serve as input for future calculations.

We conclude that the main challenge of obtaining MExFM on NiO is magnetic tip stability. Without applying a magnetic field, the magnitude of the exchange contrast on NiO using Fe-tips is much smaller $(100-400\,\mathrm{fm})$ than when applying a field of $5\,\mathrm{T}$ \cite{Kaiser2007a,Kaiser2008,Schwarz2009}. However, contrast with a similar magnitude $(1.35\,\mathrm{pm})$ can be achieved when using SmCo-tips, suggesting that the increased MAE of SmCo helps to stabilize the spin at the tip apex. The MAE of SmCo is approximately $1\,\mathrm{meV}$ per atom, almost equal to the Zeeman energy $E_Z=g \mu_B B=0.6\,\textnormal{meV}$ for a g-factor of $2.2$ for Fe and $B=5\,\textnormal{T}$ \cite{Kaiser2007a}.
Our study is a step towards a more detailed understanding of the interaction mechanism in magnetic exchange force microscopy on insulating surfaces. Based on these findings, we propose materials with high MAE to be best suited for MExFM studies. This is of particular import for the study of antiferromagnetic pinning layers in exchange bias coupled systems.

The authors thank F. Oberhuber for FIB tip etching, G. Bayreuther for discussions and the Deutsche Forschungsgemeinschaft for funding within the SFB 689.


\begin{thebibliography}{38}%
\makeatletter
\providecommand \@ifxundefined [1]{%
 \@ifx{#1\undefined}
}%
\providecommand \@ifnum [1]{%
 \ifnum #1\expandafter \@firstoftwo
 \else \expandafter \@secondoftwo
 \fi
}%
\providecommand \@ifx [1]{%
 \ifx #1\expandafter \@firstoftwo
 \else \expandafter \@secondoftwo
 \fi
}%
\providecommand \natexlab [1]{#1}%
\providecommand \enquote  [1]{``#1''}%
\providecommand \bibnamefont  [1]{#1}%
\providecommand \bibfnamefont [1]{#1}%
\providecommand \citenamefont [1]{#1}%
\providecommand \href@noop [0]{\@secondoftwo}%
\providecommand \href [0]{\begingroup \@sanitize@url \@href}%
\providecommand \@href[1]{\@@startlink{#1}\@@href}%
\providecommand \@@href[1]{\endgroup#1\@@endlink}%
\providecommand \@sanitize@url [0]{\catcode `\\12\catcode `\$12\catcode
  `\&12\catcode `\#12\catcode `\^12\catcode `\_12\catcode `\%12\relax}%
\providecommand \@@startlink[1]{}%
\providecommand \@@endlink[0]{}%
\providecommand \url  [0]{\begingroup\@sanitize@url \@url }%
\providecommand \@url [1]{\endgroup\@href {#1}{\urlprefix }}%
\providecommand \urlprefix  [0]{URL }%
\providecommand \Eprint [0]{\href }%
\providecommand \doibase [0]{http://dx.doi.org/}%
\providecommand \selectlanguage [0]{\@gobble}%
\providecommand \bibinfo  [0]{\@secondoftwo}%
\providecommand \bibfield  [0]{\@secondoftwo}%
\providecommand \translation [1]{[#1]}%
\providecommand \BibitemOpen [0]{}%
\providecommand \bibitemStop [0]{}%
\providecommand \bibitemNoStop [0]{.\EOS\space}%
\providecommand \EOS [0]{\spacefactor3000\relax}%
\providecommand \BibitemShut  [1]{\csname bibitem#1\endcsname}%
\let\auto@bib@innerbib\@empty
\bibitem [{\citenamefont {Binnig}\ \emph {et~al.}(1986)\citenamefont {Binnig},
  \citenamefont {Quate},\ and\ \citenamefont {Gerber}}]{Binnig1986}%
  \BibitemOpen
  \bibfield  {author} {\bibinfo {author} {\bibfnamefont {G.}~\bibnamefont
  {Binnig}}, \bibinfo {author} {\bibfnamefont {C.~F.}\ \bibnamefont {Quate}}, \
  and\ \bibinfo {author} {\bibfnamefont {C.}~\bibnamefont {Gerber}},\ }\href
  {\doibase 10.1103/PhysRevLett.56.930} {\bibfield  {journal} {\bibinfo
  {journal} {Physical Review Letters}\ }\textbf {\bibinfo {volume} {56}},\
  \bibinfo {pages} {930} (\bibinfo {year} {1986})}\BibitemShut {NoStop}%
\bibitem [{\citenamefont {Giessibl}(1995)}]{Giessibl1995}%
  \BibitemOpen
  \bibfield  {author} {\bibinfo {author} {\bibfnamefont {F.~J.}\ \bibnamefont
  {Giessibl}},\ }\href {\doibase 10.1126/science.267.5194.68} {\bibfield
  {journal} {\bibinfo  {journal} {Science}\ }\textbf {\bibinfo {volume}
  {267}},\ \bibinfo {pages} {68} (\bibinfo {year} {1995})}\BibitemShut
  {NoStop}%
\bibitem [{\citenamefont {Lantz}\ \emph {et~al.}(2001)\citenamefont {Lantz},
  \citenamefont {Hug}, \citenamefont {Hoffmann}, \citenamefont {van Schendel},
  \citenamefont {Kappenberger}, \citenamefont {Martin}, \citenamefont
  {Baratoff},\ and\ \citenamefont {G\"{u}ntherodt}}]{Lantz2001}%
  \BibitemOpen
  \bibfield  {author} {\bibinfo {author} {\bibfnamefont {M.~A.}\ \bibnamefont
  {Lantz}}, \bibinfo {author} {\bibfnamefont {H.~J.}\ \bibnamefont {Hug}},
  \bibinfo {author} {\bibfnamefont {R.}~\bibnamefont {Hoffmann}}, \bibinfo
  {author} {\bibfnamefont {P.~J.}\ \bibnamefont {van Schendel}}, \bibinfo
  {author} {\bibfnamefont {P.}~\bibnamefont {Kappenberger}}, \bibinfo {author}
  {\bibfnamefont {S.}~\bibnamefont {Martin}}, \bibinfo {author} {\bibfnamefont
  {A.}~\bibnamefont {Baratoff}}, \ and\ \bibinfo {author} {\bibfnamefont
  {H.~J.}\ \bibnamefont {G\"{u}ntherodt}},\ }\href {\doibase
  10.1126/science.1057824} {\bibfield  {journal} {\bibinfo  {journal}
  {Science}\ }\textbf {\bibinfo {volume} {291}},\ \bibinfo {pages} {2580}
  (\bibinfo {year} {2001})}\BibitemShut {NoStop}%
\bibitem [{\citenamefont {Martin}\ and\ \citenamefont
  {Wickramasinghe}(1987)}]{Martin1987}%
  \BibitemOpen
  \bibfield  {author} {\bibinfo {author} {\bibfnamefont {Y.}~\bibnamefont
  {Martin}}\ and\ \bibinfo {author} {\bibfnamefont {H.~K.}\ \bibnamefont
  {Wickramasinghe}},\ }\href {\doibase 10.1063/1.97800} {\bibfield  {journal}
  {\bibinfo  {journal} {Applied Physics Letters}\ }\textbf {\bibinfo {volume}
  {50}},\ \bibinfo {pages} {1455} (\bibinfo {year} {1987})}\BibitemShut
  {NoStop}%
\bibitem [{\citenamefont {Koblischka}\ \emph {et~al.}(2003)\citenamefont
  {Koblischka}, \citenamefont {Hartmann},\ and\ \citenamefont
  {Sulzbach}}]{Koblischka2003}%
  \BibitemOpen
  \bibfield  {author} {\bibinfo {author} {\bibfnamefont {M.~R.}\ \bibnamefont
  {Koblischka}}, \bibinfo {author} {\bibfnamefont {U.}~\bibnamefont
  {Hartmann}}, \ and\ \bibinfo {author} {\bibfnamefont {T.}~\bibnamefont
  {Sulzbach}},\ }\href
  {http://www.sciencedirect.com/science/article/pii/S0928493103001887}
  {\bibfield  {journal} {\bibinfo  {journal} {Materials Science and
  Engineering: C}\ }\textbf {\bibinfo {volume} {23}},\ \bibinfo {pages} {747}
  (\bibinfo {year} {2003})}\BibitemShut {NoStop}%
\bibitem [{\citenamefont {Wiesendanger}\ \emph {et~al.}(1991)\citenamefont
  {Wiesendanger}, \citenamefont {Tarrach}, \citenamefont {Wadas}, \citenamefont
  {Brodbeck}, \citenamefont {G\"{u}ntherodt}, \citenamefont {G\"{u}ntherodt},
  \citenamefont {Gambino},\ and\ \citenamefont {Ruf}}]{Wiesendanger1991}%
  \BibitemOpen
  \bibfield  {author} {\bibinfo {author} {\bibfnamefont {R.}~\bibnamefont
  {Wiesendanger}}, \bibinfo {author} {\bibfnamefont {G.}~\bibnamefont
  {Tarrach}}, \bibinfo {author} {\bibfnamefont {A.}~\bibnamefont {Wadas}},
  \bibinfo {author} {\bibfnamefont {D.}~\bibnamefont {Brodbeck}}, \bibinfo
  {author} {\bibfnamefont {H.}~\bibnamefont {G\"{u}ntherodt}}, \bibinfo
  {author} {\bibfnamefont {G.}~\bibnamefont {G\"{u}ntherodt}}, \bibinfo
  {author} {\bibfnamefont {R.}~\bibnamefont {Gambino}}, \ and\ \bibinfo
  {author} {\bibfnamefont {R.}~\bibnamefont {Ruf}},\ }\href@noop {} {\bibfield
  {journal} {\bibinfo  {journal} {Journal of Vacuum Science \& Technology B}\
  }\textbf {\bibinfo {volume} {9}},\ \bibinfo {pages} {519} (\bibinfo {year}
  {1991})}\BibitemShut {NoStop}%
\bibitem [{\citenamefont {Mukasa}\ \emph {et~al.}(1994)\citenamefont {Mukasa},
  \citenamefont {Hasegawa}, \citenamefont {Tazuke}, \citenamefont {Sueoka},
  \citenamefont {Sasaki},\ and\ \citenamefont {Hayakawa}}]{Mukasa1994a}%
  \BibitemOpen
  \bibfield  {author} {\bibinfo {author} {\bibfnamefont {K.}~\bibnamefont
  {Mukasa}}, \bibinfo {author} {\bibfnamefont {H.}~\bibnamefont {Hasegawa}},
  \bibinfo {author} {\bibfnamefont {Y.}~\bibnamefont {Tazuke}}, \bibinfo
  {author} {\bibfnamefont {K.}~\bibnamefont {Sueoka}}, \bibinfo {author}
  {\bibfnamefont {M.}~\bibnamefont {Sasaki}}, \ and\ \bibinfo {author}
  {\bibfnamefont {K.}~\bibnamefont {Hayakawa}},\ }\href {\doibase
  10.1143/JJAP.33.2692} {\bibfield  {journal} {\bibinfo  {journal} {Japanese
  Journal of Applied Physics}\ }\textbf {\bibinfo {volume} {33}},\ \bibinfo
  {pages} {2692} (\bibinfo {year} {1994})}\BibitemShut {NoStop}%
\bibitem [{\citenamefont {Ness}\ and\ \citenamefont
  {Gautier}(1995)}]{Ness1995}%
  \BibitemOpen
  \bibfield  {author} {\bibinfo {author} {\bibfnamefont {H.}~\bibnamefont
  {Ness}}\ and\ \bibinfo {author} {\bibfnamefont {F.}~\bibnamefont {Gautier}},\
  }\href {\doibase 10.1103/PhysRevB.52.7352} {\bibfield  {journal} {\bibinfo
  {journal} {Physical Review B}\ }\textbf {\bibinfo {volume} {52}},\ \bibinfo
  {pages} {7352} (\bibinfo {year} {1995})}\BibitemShut {NoStop}%
\bibitem [{\citenamefont {Nakamura}\ \emph {et~al.}(1997)\citenamefont
  {Nakamura}, \citenamefont {Hasegawa}, \citenamefont {Oguchi}, \citenamefont
  {Sueoka}, \citenamefont {Hayakawa},\ and\ \citenamefont
  {Mukasa}}]{Nakamura1997}%
  \BibitemOpen
  \bibfield  {author} {\bibinfo {author} {\bibfnamefont {K.}~\bibnamefont
  {Nakamura}}, \bibinfo {author} {\bibfnamefont {H.}~\bibnamefont {Hasegawa}},
  \bibinfo {author} {\bibfnamefont {T.}~\bibnamefont {Oguchi}}, \bibinfo
  {author} {\bibfnamefont {K.}~\bibnamefont {Sueoka}}, \bibinfo {author}
  {\bibfnamefont {K.}~\bibnamefont {Hayakawa}}, \ and\ \bibinfo {author}
  {\bibfnamefont {K.}~\bibnamefont {Mukasa}},\ }\href
  {http://arxiv.org/abs/cond-mat/9705149} {\bibfield  {journal} {\bibinfo
  {journal} {Physical Review B}\ }\textbf {\bibinfo {volume} {56}},\ \bibinfo
  {pages} {3218} (\bibinfo {year} {1997})}\BibitemShut {NoStop}%
\bibitem [{\citenamefont {Foster}\ and\ \citenamefont
  {Shluger}(2001)}]{Foster2001}%
  \BibitemOpen
  \bibfield  {author} {\bibinfo {author} {\bibfnamefont {A.}~\bibnamefont
  {Foster}}\ and\ \bibinfo {author} {\bibfnamefont {A.}~\bibnamefont
  {Shluger}},\ }\href
  {http://www.sciencedirect.com/science/article/pii/S0039602801013346}
  {\bibfield  {journal} {\bibinfo  {journal} {Surface Science}\ }\textbf
  {\bibinfo {volume} {490}},\ \bibinfo {pages} {211} (\bibinfo {year}
  {2001})}\BibitemShut {NoStop}%
\bibitem [{\citenamefont {Momida}\ and\ \citenamefont
  {Oguchi}(2005)}]{Momida2005}%
  \BibitemOpen
  \bibfield  {author} {\bibinfo {author} {\bibfnamefont {H.}~\bibnamefont
  {Momida}}\ and\ \bibinfo {author} {\bibfnamefont {T.}~\bibnamefont
  {Oguchi}},\ }\href {\doibase 10.1016/j.susc.2005.06.006} {\bibfield
  {journal} {\bibinfo  {journal} {Surface Science}\ }\textbf {\bibinfo {volume}
  {590}},\ \bibinfo {pages} {42} (\bibinfo {year} {2005})}\BibitemShut
  {NoStop}%
\bibitem [{\citenamefont {Hosoi}\ \emph {et~al.}(2000)\citenamefont {Hosoi},
  \citenamefont {Seuoka}, \citenamefont {Hayakawa},\ and\ \citenamefont
  {Mukasa}}]{Hosoi2000}%
  \BibitemOpen
  \bibfield  {author} {\bibinfo {author} {\bibfnamefont {H.}~\bibnamefont
  {Hosoi}}, \bibinfo {author} {\bibfnamefont {K.}~\bibnamefont {Seuoka}},
  \bibinfo {author} {\bibfnamefont {K.}~\bibnamefont {Hayakawa}}, \ and\
  \bibinfo {author} {\bibfnamefont {K.}~\bibnamefont {Mukasa}},\ }\href
  {\doibase 10.1016/S0169-4332(99)00529-2} {\bibfield  {journal} {\bibinfo
  {journal} {Applied Surface Science}\ }\textbf {\bibinfo {volume} {157}},\
  \bibinfo {pages} {218} (\bibinfo {year} {2000})}\BibitemShut {NoStop}%
\bibitem [{\citenamefont {Allers}\ \emph {et~al.}(2001)\citenamefont {Allers},
  \citenamefont {Langkat},\ and\ \citenamefont {Wiesendanger}}]{Allers2001}%
  \BibitemOpen
  \bibfield  {author} {\bibinfo {author} {\bibfnamefont {W.}~\bibnamefont
  {Allers}}, \bibinfo {author} {\bibfnamefont {S.}~\bibnamefont {Langkat}}, \
  and\ \bibinfo {author} {\bibfnamefont {R.}~\bibnamefont {Wiesendanger}},\
  }\href {\doibase 10.1007/s003390100731} {\bibfield  {journal} {\bibinfo
  {journal} {Applied Physics}\ }\textbf {\bibinfo {volume} {72}},\ \bibinfo
  {pages} {S27} (\bibinfo {year} {2001})}\BibitemShut {NoStop}%
\bibitem [{\citenamefont {Hosoi}\ \emph {et~al.}(2004)\citenamefont {Hosoi},
  \citenamefont {Sueoka},\ and\ \citenamefont {Mukasa}}]{Hosoi2004}%
  \BibitemOpen
  \bibfield  {author} {\bibinfo {author} {\bibfnamefont {H.}~\bibnamefont
  {Hosoi}}, \bibinfo {author} {\bibfnamefont {K.}~\bibnamefont {Sueoka}}, \
  and\ \bibinfo {author} {\bibfnamefont {K.}~\bibnamefont {Mukasa}},\ }\href
  {\doibase 10.1088/0957-4484/15/5/018} {\bibfield  {journal} {\bibinfo
  {journal} {Nanotechnology}\ }\textbf {\bibinfo {volume} {15}},\ \bibinfo
  {pages} {505} (\bibinfo {year} {2004})}\BibitemShut {NoStop}%
\bibitem [{\citenamefont {Schmid}\ \emph {et~al.}(2008)\citenamefont {Schmid},
  \citenamefont {Mannhart},\ and\ \citenamefont {Giessibl}}]{Schmid2008}%
  \BibitemOpen
  \bibfield  {author} {\bibinfo {author} {\bibfnamefont {M.}~\bibnamefont
  {Schmid}}, \bibinfo {author} {\bibfnamefont {J.}~\bibnamefont {Mannhart}}, \
  and\ \bibinfo {author} {\bibfnamefont {F.~J.}\ \bibnamefont {Giessibl}},\
  }\href {\doibase 10.1103/PhysRevB.77.045402} {\bibfield  {journal} {\bibinfo
  {journal} {Physical Review B}\ }\textbf {\bibinfo {volume} {77}},\ \bibinfo
  {pages} {045402} (\bibinfo {year} {2008})}\BibitemShut {NoStop}%
  \bibitem [{\citenamefont {Kaiser}\ \emph {et~al.}(2007)\citenamefont {Kaiser},
  \citenamefont {Schwarz},\ and\ \citenamefont {Wiesendanger}}]{Kaiser2007a}%
  \BibitemOpen
  \bibfield  {author} {\bibinfo {author} {\bibfnamefont {U.}~\bibnamefont
  {Kaiser}}, \bibinfo {author} {\bibfnamefont {A.}~\bibnamefont {Schwarz}}, \
  and\ \bibinfo {author} {\bibfnamefont {R.}~\bibnamefont {Wiesendanger}},\
  }\href {\doibase 10.1038/nature05617} {\bibfield  {journal} {\bibinfo
  {journal} {Nature}\ }\textbf {\bibinfo {volume} {446}},\ \bibinfo {pages}
  {522} (\bibinfo {year} {2007})}\BibitemShut {NoStop}%
\bibitem [{\citenamefont {Kaiser}\ \emph {et~al.}(2008)\citenamefont {Kaiser},
  \citenamefont {Schwarz},\ and\ \citenamefont {Wiesendanger}}]{Kaiser2008}%
  \BibitemOpen
  \bibfield  {author} {\bibinfo {author} {\bibfnamefont {U.}~\bibnamefont
  {Kaiser}}, \bibinfo {author} {\bibfnamefont {A.}~\bibnamefont {Schwarz}}, \
  and\ \bibinfo {author} {\bibfnamefont {R.}~\bibnamefont {Wiesendanger}},\
  }\href {\doibase 10.1103/PhysRevB.78.104418} {\bibfield  {journal} {\bibinfo
  {journal} {Physical Review B}\ }\textbf {\bibinfo {volume} {78}},\ \bibinfo
  {pages} {104418} (\bibinfo {year} {2008})}\BibitemShut {NoStop}%
\bibitem [{\citenamefont {Schwarz}\ \emph {et~al.}(2009)\citenamefont
  {Schwarz}, \citenamefont {Kaiser},\ and\ \citenamefont
  {Wiesendanger}}]{Schwarz2009}%
  \BibitemOpen
  \bibfield  {author} {\bibinfo {author} {\bibfnamefont {A.}~\bibnamefont
  {Schwarz}}, \bibinfo {author} {\bibfnamefont {U.}~\bibnamefont {Kaiser}}, \
  and\ \bibinfo {author} {\bibfnamefont {R.}~\bibnamefont {Wiesendanger}},\
  }\href {\doibase 10.1088/0957-4484/20/26/264017} {\bibfield  {journal}
  {\bibinfo  {journal} {Nanotechnology}\ }\textbf {\bibinfo {volume} {20}},\
  \bibinfo {pages} {264017} (\bibinfo {year} {2009})}\BibitemShut {NoStop}%
\bibitem [{\citenamefont {Givord}\ \emph {et~al.}(1979)\citenamefont {Givord},
  \citenamefont {Laforest}, \citenamefont {Schweizer},\ and\ \citenamefont
  {Tasset}}]{Givord1979}%
  \BibitemOpen
  \bibfield  {author} {\bibinfo {author} {\bibfnamefont {D.}~\bibnamefont
  {Givord}}, \bibinfo {author} {\bibfnamefont {J.}~\bibnamefont {Laforest}},
  \bibinfo {author} {\bibfnamefont {J.}~\bibnamefont {Schweizer}}, \ and\
  \bibinfo {author} {\bibfnamefont {F.}~\bibnamefont {Tasset}},\ }\href
  {\doibase 10.1063/1.327141} {\bibfield  {journal} {\bibinfo  {journal}
  {Journal of Applied Physics}\ }\textbf {\bibinfo {volume} {50}},\ \bibinfo
  {pages} {2008} (\bibinfo {year} {1979})}\BibitemShut {NoStop}%
\bibitem [{\citenamefont {Schmidt}\ \emph {et~al.}(2011)\citenamefont
  {Schmidt}, \citenamefont {Lazo}, \citenamefont {Kaiser}, \citenamefont
  {Schwarz}, \citenamefont {Heinze},\ and\ \citenamefont
  {Wiesendanger}}]{Schmidt2011}%
  \BibitemOpen
  \bibfield  {author} {\bibinfo {author} {\bibfnamefont {R.}~\bibnamefont
  {Schmidt}}, \bibinfo {author} {\bibfnamefont {C.}~\bibnamefont {Lazo}},
  \bibinfo {author} {\bibfnamefont {U.}~\bibnamefont {Kaiser}}, \bibinfo
  {author} {\bibfnamefont {A.}~\bibnamefont {Schwarz}}, \bibinfo {author}
  {\bibfnamefont {S.}~\bibnamefont {Heinze}}, \ and\ \bibinfo {author}
  {\bibfnamefont {R.}~\bibnamefont {Wiesendanger}},\ }\href {\doibase
  10.1103/PhysRevLett.106.257202} {\bibfield  {journal} {\bibinfo  {journal}
  {Physical Review Letters}\ }\textbf {\bibinfo {volume} {106}},\ \bibinfo
  {pages} {257202} (\bibinfo {year} {2011})}\BibitemShut {NoStop}%
\bibitem [{\citenamefont {Horcas}\ \emph {et~al.}(2007)\citenamefont {Horcas},
  \citenamefont {Fern\'{a}ndez}, \citenamefont {G\'{o}mez-Rodr\'{\i}guez},
  \citenamefont {Colchero}, \citenamefont {G\'{o}mez-Herrero},\ and\
  \citenamefont {Baro}}]{Horcas2007}%
  \BibitemOpen
  \bibfield  {author} {\bibinfo {author} {\bibfnamefont {I.}~\bibnamefont
  {Horcas}}, \bibinfo {author} {\bibfnamefont {R.}~\bibnamefont
  {Fern\'{a}ndez}}, \bibinfo {author} {\bibfnamefont {J.~M.}\ \bibnamefont
  {G\'{o}mez-Rodr\'{\i}guez}}, \bibinfo {author} {\bibfnamefont
  {J.}~\bibnamefont {Colchero}}, \bibinfo {author} {\bibfnamefont
  {J.}~\bibnamefont {G\'{o}mez-Herrero}}, \ and\ \bibinfo {author}
  {\bibfnamefont {A.~M.}\ \bibnamefont {Baro}},\ }\href {\doibase
  10.1063/1.2432410} {\bibfield  {journal} {\bibinfo  {journal} {Review of
  Scientific Instruments}\ }\textbf {\bibinfo {volume} {78}},\ \bibinfo {pages}
  {013705} (\bibinfo {year} {2007})}\BibitemShut {NoStop}%
\bibitem [{\citenamefont {Albrecht}\ \emph {et~al.}(1991)\citenamefont
  {Albrecht}, \citenamefont {Gr\"{u}tter}, \citenamefont {Horne},\ and\
  \citenamefont {Rugar}}]{Albrecht1991}%
  \BibitemOpen
  \bibfield  {author} {\bibinfo {author} {\bibfnamefont {T.~R.}\ \bibnamefont
  {Albrecht}}, \bibinfo {author} {\bibfnamefont {P.}~\bibnamefont
  {Gr\"{u}tter}}, \bibinfo {author} {\bibfnamefont {D.}~\bibnamefont {Horne}},
  \ and\ \bibinfo {author} {\bibfnamefont {D.}~\bibnamefont {Rugar}},\
  }\href@noop {} {\bibfield  {journal} {\bibinfo  {journal} {Journal of Applied
  Physics}\ }\textbf {\bibinfo {volume} {69}},\ \bibinfo {pages} {668} (\bibinfo
  {year} {1991})}\BibitemShut {NoStop}%
\bibitem [{\citenamefont {Giessibl}(1997)}]{Giessibl1997}%
  \BibitemOpen
  \bibfield  {author} {\bibinfo {author} {\bibfnamefont {F.~J.}\ \bibnamefont
  {Giessibl}},\ }\href@noop {} {\bibfield  {journal} {\bibinfo  {journal}
  {Physical Review B}\ }\textbf {\bibinfo {volume} {56}},\ \bibinfo {pages}
  {16010} (\bibinfo {year} {1997})}\BibitemShut {NoStop}%
\bibitem [{\citenamefont {Sader}\ and\ \citenamefont
  {Jarvis}(2004)}]{Sader2004}%
  \BibitemOpen
  \bibfield  {author} {\bibinfo {author} {\bibfnamefont {J.~E.}\ \bibnamefont
  {Sader}}\ and\ \bibinfo {author} {\bibfnamefont {S.~P.}\ \bibnamefont
  {Jarvis}},\ }\href {\doibase 10.1063/1.1667267} {\bibfield  {journal}
  {\bibinfo  {journal} {Applied Physics Letters}\ }\textbf {\bibinfo {volume}
  {84}},\ \bibinfo {pages} {1801} (\bibinfo {year} {2004})}\BibitemShut
  {NoStop}%
\bibitem [{\citenamefont {Giessibl}\ \emph {et~al.}(1999)\citenamefont
  {Giessibl}, \citenamefont {Bielefeldt}, \citenamefont {Hembacher},\ and\
  \citenamefont {Mannhart}}]{Giessibl1999}%
  \BibitemOpen
  \bibfield  {author} {\bibinfo {author} {\bibfnamefont {F.~J.}\ \bibnamefont
  {Giessibl}}, \bibinfo {author} {\bibfnamefont {H.}~\bibnamefont
  {Bielefeldt}}, \bibinfo {author} {\bibfnamefont {S.}~\bibnamefont
  {Hembacher}}, \ and\ \bibinfo {author} {\bibfnamefont {J.}~\bibnamefont
  {Mannhart}},\ }\href@noop {}  {\bibfield  {journal}
  {\bibinfo  {journal} {Applied Surface Science}\ }\textbf {\bibinfo {volume}
  {140}},\ \bibinfo {pages} {352} (\bibinfo {year} {1999})}\BibitemShut  {NoStop}%
\bibitem [{\citenamefont {Ternes}\ \emph {et~al.}(2008)\citenamefont {Ternes},
  \citenamefont {Lutz}, \citenamefont {Hirjibehedin}, \citenamefont
  {Giessibl},\ and\ \citenamefont {Heinrich}}]{Ternes2008}%
  \BibitemOpen
  \bibfield  {author} {\bibinfo {author} {\bibfnamefont {M.}~\bibnamefont
  {Ternes}}, \bibinfo {author} {\bibfnamefont {C.~P.}\ \bibnamefont {Lutz}},
  \bibinfo {author} {\bibfnamefont {C.~F.}\ \bibnamefont {Hirjibehedin}},
  \bibinfo {author} {\bibfnamefont {F.~J.}\ \bibnamefont {Giessibl}}, \ and\
  \bibinfo {author} {\bibfnamefont {A.~J.}\ \bibnamefont {Heinrich}},\
  }\href@noop {} {\bibfield  {journal} {\bibinfo  {journal}   
{Science}\ }\textbf {\bibinfo {volume} {319}},\ \bibinfo {pages} {1066}
  (\bibinfo {year} {2008})}\BibitemShut {NoStop}%
\bibitem [{\citenamefont {Gross}\ \emph {et~al.}(2009)\citenamefont {Gross},
  \citenamefont {Mohn}, \citenamefont {Moll}, \citenamefont {Liljeroth},\ and\
  \citenamefont {Meyer}}]{Gross2009}%
  \BibitemOpen
  \bibfield  {author} {\bibinfo {author} {\bibfnamefont {L.}~\bibnamefont
  {Gross}}, \bibinfo {author} {\bibfnamefont {F.}~\bibnamefont {Mohn}},
  \bibinfo {author} {\bibfnamefont {N.}~\bibnamefont {Moll}}, \bibinfo {author}
  {\bibfnamefont {P.}~\bibnamefont {Liljeroth}}, \ and\ \bibinfo {author}
  {\bibfnamefont {G.}~\bibnamefont {Meyer}},\ }\href {\doibase
  10.1126/science.1176210} {\bibfield  {journal} {\bibinfo  {journal}
  {Science}\ }\textbf {\bibinfo {volume} {325}},\ \bibinfo {pages} {1110}
  (\bibinfo {year} {2009})}\BibitemShut {NoStop}%
\bibitem [{\citenamefont {Giessibl}(1998)}]{Giessibl1998}%
  \BibitemOpen
  \bibfield  {author} {\bibinfo {author} {\bibfnamefont {F.~J.}\ \bibnamefont
  {Giessibl}},\ }\href@noop {} {\bibfield  {journal} {\bibinfo  {journal}
  {Applied Physics Letters}\ }\textbf {\bibinfo {volume} {73}},\ \bibinfo
  {pages} {3956} (\bibinfo {year} {1998})}\BibitemShut {NoStop}%
\bibitem [{\citenamefont {Herz}\ \emph {et~al.}(2003)\citenamefont {Herz},
  \citenamefont {Giessibl},\ and\ \citenamefont {Mannhart}}]{Herz2003}%
  \BibitemOpen
  \bibfield  {author} {\bibinfo {author} {\bibfnamefont {M.}~\bibnamefont
  {Herz}}, \bibinfo {author} {\bibfnamefont {F.~J.}\ \bibnamefont {Giessibl}},
  \ and\ \bibinfo {author} {\bibfnamefont {J.}~\bibnamefont {Mannhart}},\
  }\href {\doibase 10.1103/PhysRevB.68.045301} {\bibfield  {journal} {\bibinfo
  {journal} {Physical Review B}\ }\textbf {\bibinfo {volume} {68}},\ \bibinfo
  {pages} {045301} (\bibinfo {year} {2003})}\BibitemShut {NoStop}%
\bibitem [{\citenamefont {M\"{u}ller}\ \emph {et~al.}(1965)\citenamefont
  {M\"{u}ller}, \citenamefont {Nakamura}, \citenamefont {Nishikawa},\ and\
  \citenamefont {McLane}}]{Muller}%
  \BibitemOpen
  \bibfield  {author} {\bibinfo {author} {\bibfnamefont {E.~W.}\ \bibnamefont
  {M\"{u}ller}}, \bibinfo {author} {\bibfnamefont {S.}~\bibnamefont
  {Nakamura}}, \bibinfo {author} {\bibfnamefont {O.}~\bibnamefont {Nishikawa}},
  \ and\ \bibinfo {author} {\bibfnamefont {S.~B.}\ \bibnamefont {McLane}},\
  }\href {file:///C:/Users/Flo/AppData/Local/Mendeley Ltd/Mendeley
  Desktop/Downloaded/Unknown - Unknown - gas-surface interactions and field-ion
  microscopy of nonrefractory metals.html} {\bibfield  {journal} {\bibinfo
  {journal} {Journal of Applied Physics}\ }\textbf {\bibinfo {volume} {36}},\
  \bibinfo {pages} {2496} (\bibinfo {year} {1965})}\BibitemShut {NoStop}%
\bibitem [{\citenamefont {Hillebrecht}\ \emph {et~al.}(2001)\citenamefont
  {Hillebrecht}, \citenamefont {Ohldag}, \citenamefont {Weber}, \citenamefont
  {Bethke}, \citenamefont {Mick}, \citenamefont {Weiss},\ and\ \citenamefont
  {Bahrdt}}]{Hillebrecht2001}%
  \BibitemOpen
  \bibfield  {author} {\bibinfo {author} {\bibfnamefont {F.~U.}\ \bibnamefont
  {Hillebrecht}}, \bibinfo {author} {\bibfnamefont {H.}~\bibnamefont {Ohldag}},
  \bibinfo {author} {\bibfnamefont {N.~B.}\ \bibnamefont {Weber}}, \bibinfo
  {author} {\bibfnamefont {C.}~\bibnamefont {Bethke}}, \bibinfo {author}
  {\bibfnamefont {U.}~\bibnamefont {Mick}}, \bibinfo {author} {\bibfnamefont
  {M.}~\bibnamefont {Weiss}}, \ and\ \bibinfo {author} {\bibfnamefont
  {J.}~\bibnamefont {Bahrdt}},\ }\href {\doibase 10.1103/PhysRevLett.86.3419}
  {\bibfield  {journal} {\bibinfo  {journal} {Physical Review Letters}\
  }\textbf {\bibinfo {volume} {86}},\ \bibinfo {pages} {3419} (\bibinfo {year}
  {2001})}\BibitemShut {NoStop}%
  \bibitem{supp}
   See Supplemental Materials at [URL will be inserted by publisher] for detailed information on initial results with uncharacterized Fe-tips, tip characterization by the CO-method, conversion of frequency shift data to topographic data and more experimental MExFM data.
\bibitem [{\citenamefont {{Cullity, B.D.; Graham}}(2008)}]{Cullity}%
  \BibitemOpen
  \bibfield  {author} {\bibinfo {author} {\bibfnamefont {C.}~\bibnamefont
  {{Cullity, B.D.; Graham}}},\ }\href@noop {} {\emph {\bibinfo {title}
  {{Introduction to Magnetic Materials}}}}\ (\bibinfo  {publisher} {Wiley-IEEE
  Press},\ \bibinfo {year} {2008})\ p.\ \bibinfo {pages} {568}\BibitemShut
  {NoStop}%
\bibitem [{\citenamefont {Welker}\ and\ \citenamefont
  {Giessibl}(2012)}]{Welker2012}%
  \BibitemOpen
  \bibfield  {author} {\bibinfo {author} {\bibfnamefont {J.}~\bibnamefont
  {Welker}}\ and\ \bibinfo {author} {\bibfnamefont {F.~J.}\ \bibnamefont
  {Giessibl}},\ }\href {\doibase 10.1126/science.1219850} {\bibfield  {journal}
  {\bibinfo  {journal} {Science}\ }\textbf {\bibinfo {volume} {336}},\ \bibinfo
  {pages} {444} (\bibinfo {year} {2012})}\BibitemShut {NoStop}%
\bibitem [{\citenamefont {Teobaldi}\ \emph {et~al.}(2011)\citenamefont
  {Teobaldi}, \citenamefont {L\"{a}mmle}, \citenamefont {Trevethan},
  \citenamefont {Watkins}, \citenamefont {Schwarz}, \citenamefont
  {Wiesendanger},\ and\ \citenamefont {Shluger}}]{Teobaldi2011}%
  \BibitemOpen
  \bibfield  {author} {\bibinfo {author} {\bibfnamefont {G.}~\bibnamefont
  {Teobaldi}}, \bibinfo {author} {\bibfnamefont {K.}~\bibnamefont
  {L\"{a}mmle}}, \bibinfo {author} {\bibfnamefont {T.}~\bibnamefont
  {Trevethan}}, \bibinfo {author} {\bibfnamefont {M.}~\bibnamefont {Watkins}},
  \bibinfo {author} {\bibfnamefont {A.}~\bibnamefont {Schwarz}}, \bibinfo
  {author} {\bibfnamefont {R.}~\bibnamefont {Wiesendanger}}, \ and\ \bibinfo
  {author} {\bibfnamefont {A.}~\bibnamefont {Shluger}},\ }\href {\doibase
  10.1103/PhysRevLett.106.216102} {\bibfield  {journal} {\bibinfo  {journal}
  {Physical Review Letters}\ }\textbf {\bibinfo {volume} {106}},\
  \bibinfo
  {pages} {216102} (\bibinfo
  {year} {2011})}\BibitemShut {NoStop}%
\bibitem [{\citenamefont {Weller}\ \emph {et~al.}(2000)\citenamefont {Weller},
  \citenamefont {Moser}, \citenamefont {Folks}, \citenamefont {Best},
  \citenamefont {Toney}, \citenamefont {Schwickert}, \citenamefont {Thiele},\
  and\ \citenamefont {Doerner}}]{Weller2000}%
  \BibitemOpen
  \bibfield  {author} {\bibinfo {author} {\bibfnamefont {D.}~\bibnamefont
  {Weller}}, \bibinfo {author} {\bibfnamefont {A.}~\bibnamefont {Moser}},
  \bibinfo {author} {\bibfnamefont {L.}~\bibnamefont {Folks}}, \bibinfo
  {author} {\bibfnamefont {M.}~\bibnamefont {Best}}, \bibinfo {author}
  {\bibfnamefont {M.}~\bibnamefont {Toney}}, \bibinfo {author} {\bibfnamefont
  {M.}~\bibnamefont {Schwickert}}, \bibinfo {author} {\bibfnamefont {J.-U.}\
  \bibnamefont {Thiele}}, \ and\ \bibinfo {author} {\bibfnamefont
  {M.}~\bibnamefont {Doerner}},\ }\href {\doibase 10.1109/20.824418} {\bibfield
   {journal} {\bibinfo  {journal} {IEEE Transactions on Magnetics}\ }\textbf
  {\bibinfo {volume} {36}},\ \bibinfo {pages} {10} (\bibinfo {year}
  {2000})}\BibitemShut {NoStop}%
\bibitem [{\citenamefont {Burkert}\ \emph {et~al.}(2004)\citenamefont
  {Burkert}, \citenamefont {Nordstr\"{o}m}, \citenamefont {Eriksson},\ and\
  \citenamefont {Heinonen}}]{Burkert2004}%
  \BibitemOpen
  \bibfield  {author} {\bibinfo {author} {\bibfnamefont {T.}~\bibnamefont
  {Burkert}}, \bibinfo {author} {\bibfnamefont {L.}~\bibnamefont
  {Nordstr\"{o}m}}, \bibinfo {author} {\bibfnamefont {O.}~\bibnamefont
  {Eriksson}}, \ and\ \bibinfo {author} {\bibfnamefont {O.}~\bibnamefont
  {Heinonen}},\ }\href {\doibase 10.1103/PhysRevLett.93.027203} {\bibfield
  {journal} {\bibinfo  {journal} {Physical Review Letters}\ }\textbf {\bibinfo
  {volume} {93}},\ \bibinfo {pages} {027203} (\bibinfo {year}
  {2004})}\BibitemShut {NoStop}%
\bibitem [{\citenamefont {K\"{o}dderitzsch}\ \emph {et~al.}(2002)\citenamefont
  {K\"{o}dderitzsch}, \citenamefont {Hergert}, \citenamefont {Temmerman},
  \citenamefont {Szotek}, \citenamefont {Ernst},\ and\ \citenamefont
  {Winter}}]{Ködd2002}%
  \BibitemOpen
  \bibfield  {author} {\bibinfo {author} {\bibfnamefont {D.}~\bibnamefont
  {K\"{o}dderitzsch}}, \bibinfo {author} {\bibfnamefont {W.}~\bibnamefont
  {Hergert}}, \bibinfo {author} {\bibfnamefont {W.}~\bibnamefont {Temmerman}},
  \bibinfo {author} {\bibfnamefont {Z.}~\bibnamefont {Szotek}}, \bibinfo
  {author} {\bibfnamefont {A.}~\bibnamefont {Ernst}}, \ and\ \bibinfo {author}
  {\bibfnamefont {H.}~\bibnamefont {Winter}},\ }\href {\doibase
  10.1103/PhysRevB.66.064434} {\bibfield  {journal} {\bibinfo  {journal}
  {Physical Review B}\ }\textbf {\bibinfo {volume} {66}},\ \bibinfo {pages} {064434}
  (\bibinfo {year} {2002})}\BibitemShut {NoStop}%
\end{thebibliography}
\end{document}